\documentclass[fleqn,10pt]{wlscirep}
\usepackage[utf8]{inputenc}
\usepackage[T1]{fontenc}
\usepackage{graphicx}
\usepackage{adjustbox}
\usepackage{soul}

\title{COVID-Net S: Towards computer-aided severity assessment via training and validation of deep neural networks for geographic extent and opacity extent scoring of chest X-rays for SARS-CoV-2 lung disease severity}

\author[1,2*]{A.~Wong}
\author[1,2*]{Z.~Q.~Lin}
\author[1,2]{L.~Wang}
\author[2]{A.~G.~Chung}
\author[3]{B.~Shen}
\author[3]{A.~Abbasi}
\author[3]{M.~Hoshmand-Kochi}
\author[4]{T.~Q.~Duong}
\affil[1]{University of Waterloo, Systems Design Engineering, Waterloo, Canada}
\affil[2]{DarwinAI Corp., Waterloo, Canada}
\affil[3]{Stony Brook School of Medicine, Department of Radiology, Stony Brook, NY}
\affil[4]{Montefiore Medical Center and Albert Einstein College of Medicine, Department of Radiology, Bronx, NY}
\affil[*]{a28wong@uwaterloo.ca,zhong.q.lin@uwaterloo.ca}

\begin{abstract}
\textbf{Background:}
A critical step in effective care and treatment planning for severe acute respiratory syndrome coronavirus 2 (SARS-CoV-2), the cause for the coronavirus disease 2019 (COVID-19) pandemic, is the assessment of the severity of disease progression. Chest x-rays (CXRs) are often used to assess SARS-CoV-2 severity, with two important assessment metrics being extent of lung involvement and degree of opacity.
In this proof-of-concept study, we assess the feasibility of computer-aided scoring of CXRs of SARS-CoV-2 lung disease severity using a deep learning system.

\textbf{Materials and Methods:}
Data consisted of 396 CXRs from SARS-CoV-2 positive patient cases~\cite{Cohen, Figure1, actualmed}.  Geographic extent and opacity extent were scored by two board-certified expert chest radiologists (with 20+ years of experience) and a 2nd-year radiology resident. The deep neural networks used in this study, which we name \textbf{COVID-Net S}, are based on a COVID-Net network architecture. 100 versions of the network were independently learned (50 to perform geographic extent scoring and 50 to perform opacity extent scoring) using random subsets of CXRs from the study, and we evaluated the networks using stratified Monte Carlo cross-validation experiments.

\textbf{Findings:}
The COVID-Net S deep neural networks yielded R$^2$ of 0.664 $\pm$ 0.032 and 0.635 $\pm$ 0.044 between predicted scores and radiologist scores for geographic extent and opacity extent, respectively, in stratified Monte Carlo cross-validation experiments. The best performing COVID-Net S networks achieved R$^2$ of 0.739 and 0.741 between predicted scores and radiologist scores for geographic extent and opacity extent, respectively.

\textbf{Interpretation:}
The results are promising and suggest that the use of deep neural networks on CXRs could be an effective tool for computer-aided assessment of SARS-CoV-2 lung disease severity, although additional studies are needed before adoption for routine clinical use.
\end{abstract}
\begin{document}

\flushbottom
\maketitle

\thispagestyle{empty}

\section*{Introduction}
As the coronavirus disease 2019 (COVID-19) pandemic, caused by severe acute respiratory syndrome coronavirus 2 (SARS-CoV-2), continues around the world, radiology has seen growing importance in providing clinical insights for aiding the diagnosis, treatment, and management of the disease.  Much of early literature have focused on imaging features presented in computed tomography (CT) scans of SARS-CoV-2 positive patients given its use in China during the earlier stages of the global pandemic~\cite{Zhou,Chung,Ai,Fang,Shi,Gunraj2020,gunraj2021covidnet}; however, the low availability of CT scanners in many parts of the world due to its high costs, the high risk of SARS-CoV-2 transmission during patient transport to/from CT imaging suites, and long decontamination times between scans have limited the use of CT scans for SARS-CoV-2 diagnosis and treatment planning.  A number of recent studies have illustrated the growing interest and usage of chest x-ray (CXR) imaging around the world~\cite{RSNA,Jacobi,Wong,Warren,Toussie,Huang,Guan,Mao}, with some studies foreseeing a greater reliance on portable CXR~\cite{Jacobi} and the high value of portable CXR for critically ill patients~\cite{Wu}. Compared to CT scanners, CXR imaging systems are widely available around the world due to their relatively low cost, and have comparatively faster decontamination times; in addition, the existence of portable CXR units means that imaging can occur within an isolation room and, thus, greatly reduce transmission risk~\cite{Jacobi,RSNA,CSTR}.  Furthermore, CXR imaging is frequently performed for patients with respiratory complaints as part of standard procedure~\cite{BSTI}, and have been shown to give valuable insights on disease progression~\cite{Wong}.  In the context of detecting SARS-CoV-2, CXR imaging can also be useful in situations where patients with initial negative reverse transcription–polymerase chain reaction (RT-PCR) results, the current gold standard for viral testing, revisit the emergency department with worsening symptoms~\cite{CSTR}.

Several studies have investigated imaging features presented in CXR images of SARS-CoV-2 positive patients~\cite{Huang, Guan, Kong}, with commonly found features being bilateral abnormalities, ground-glass opacity, and interstitial abnormalities.  Leveraging the presence of these imaging features in combination with the ability to observe their progression and extent over the duration of disease onset, an important role that CXR assessment has in aiding with disease treatment and management is in determining the severity of a patient's condition.  As such, a number of recent studies have focused on severity scoring~\cite{Wong,Warren,Toussie}, where the goal is to quantify SARS-CoV-2 lung disease severity. Disease severity scoring can help with determining the best course of treatment and management given a SARS-CoV-2 case (e.g., at-home quarantine, oxygen therapy, ventilation, etc.), allowing for the individualized treatment of each patient.

We hypothesise that deep learning could potentially be a valuable tool for enabling computer-aided severity scoring of SARS-CoV-2 lung severity using CXRs of SARS-CoV-2 positive patients.  Using CXR training data acquired from a global pool of SARS-CoV-2 positive patients, deep neural networks can learn to identify the important imaging features within a CXR image indicative of SARS-CoV-2, and output scores for quantifying the severity of a patient's disease progression.  In this study, we assess the feasibility of computer-aided severity scoring of SARS-CoV-2 lung severity using deep learning by developing, training, and validating 100 versions of a deep neural network we name \textbf{COVID-Net S} (50 for performing geographic extent scoring and 50 for performing opacity scoring) using stratified Monte Carlo cross-validation experiments on data consisting of 396 CXRs from positive patient cases. Two board-certified chest radiologists and a radiology resident assess the results achieved by the deep neural networks.

\section*{Materials and Methods}

\subsection*{Data preparation and radiological scoring}
The primary goal of this study is to assess the feasibility of computer-aided severity scoring of SARS-CoV-2 using deep learning.  To this end, we develop and evaluate deep neural networks that can score CXRs of patients with SARS-CoV-2.  Data consisted of CXR data pertaining to SARS-CoV-2 positive cases~\cite{Cohen, Figure1, actualmed}.  In this study specifically, the 396 CXRs from the studies used here represent a patient population of 267 patients between 12 and 88 years old around the world.  The CXR data were acquired using a range of X-ray imaging equipment types and acquisition protocols that are representative of routine imaging practice (including supine and upright, posterioranterior and anteriorposterior).

Radiological scoring was performed by two board-certified chest radiologists with 20+ years of experience (A.A. and M.H.) and a 2nd-year radiology resident (B.S.) to stage SARS-CoV-2 disease severity using a score system adapted from Wong~\textit{et al.}~\cite{Wong} and Warren~\textit{et al.}~\cite{Warren}. The two assessment metrics scored in the radiological scoring are geographic extent and opacity extent.  More specifically, for geographic extent, the extent of lung involvement by ground glass opacity or consolidation of each lung (with the right and left lung scored separately) is scored as: 0 = no involvement; 1 = <25\%; 2 = 25-50\%; 3 = 50-75\%; 4 = >75\% involvement.  The scores are then added together, and the total geographic extent score ranges from 0 to 8 (right + left lung).  For opacity extent, the degree of opacity was similarly scored for the right and left lung separately as: 0 = no opacity; 1 = ground glass opacity; 2 = mix of consolidation and ground glass opacity (less than 50\% consolidation); 3 = mix of consolidation and ground glass opacity (more than 50\% consolidation); 4 = complete white-out.  The scores are similarly added together, and the total opacity extent score ranges from 0 to 8 (right + left lung).  The mean scores are then calculated across the radiologists and used to train the deep neural networks.  The inter-reader agreement assessed by intra-class correlation coefficient was 0.92 (95\% CI: 0.91-0.93) for the geographic extent scores, and 0.87 (95\% CI: 0.85-0.89) for the opacity extent scores.

After radiological scoring, all CXR data used in this study underwent data processing to facilitate the training of deep neural networks. To discourage the deep neural networks from learning irrelevant visual cues when making severity scoring predictions, the top 8\% of the CXR data were cropped to remove boundary artifacts and embedded metadata outside of the patient region of interest. Furthermore, all CXR data were resized to the same data dimensions to enable training of the deep neural networks in this study. Finally, the geographic extent scores (with a dynamic range of 0 to 8) and opacity extent scores (with a dynamic range of 0 to 8) were re-mapped to a unified dynamic range from 0 to 1.

\subsection*{Model development}

\begin{figure*}[t]
\centering
  \includegraphics[width=1\textwidth]{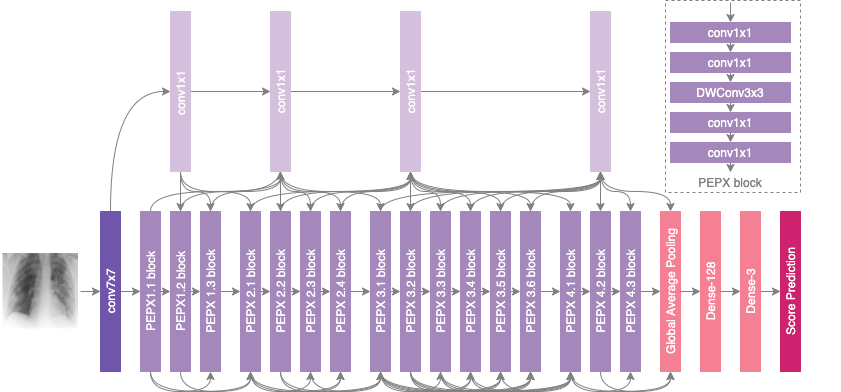}
  \caption{Flowchart of the overall architecture of the COVID-Net S deep neural networks for predicting SARS-CoV-2 severity scores.}
  \label{fig:arch}
\end{figure*}

The development of the deep neural network architecture for computer-aided severity scoring is important as it dictates the sequence of mathematical operations that maps the input CXR data to the predicted severity scores (e.g., geographic extent score and opacity extent score).  Specifically, the architecture of the deep neural network will affect the efficiency and effectiveness with which it is able to learn the underlying parameters and operations in this complex, hierarchical mapping.  In this study, the architecture of the deep neural networks used to evaluate the feasibility of computer-aided severity scoring of SARS-CoV-2 lung disease severity, which we name COVID-Net S, is based on the COVID-Net deep neural network architecture~\cite{COVIDNet}, which was found to achieve state-of-the-art performance in SARS-CoV-2 detection. The last layers of the COVID-Net architecture are replaced with a set of new layers (a 128 neuron dense layer, a 3 neuron dense layer, and a single output score prediction layer) to enable the prediction of severity scores corresponding to scores within the dynamic range of 0 to 1.  These scores can be mapped back to the original dynamic ranges of geographic extent score and opacity extent score used during radiological scoring. Figure 1 presents an overview of this COVID-Net S network architecture and is publicly available for open access at \url{https://github.com/lindawangg/COVID-Net}. The network architecture consists of projection-expansion-projection design patterns for high representational capacity while maintaining computational efficiency, selective long-range connectivity to improve learning efficiency, and high architectural diversity.

To improve the performance of the deep neural networks, a technique known as transfer learning~\cite{Pan} is used to initialize the deep neural network parameters in this study using the parameters from deep neural networks trained on COVIDx, a dataset introduced in the Wang study~\cite{COVIDNet} containing 13,975 CXR images across 13,870 patient cases consisting of healthy patients and patients with different forms of pneumonia (e.g., viral, bacterial, etc.).  Statistical distribution details of COVIDx can be found in the Wang study~\cite{COVIDNet}.  We also leverage data augmentation~\cite{Perez} in this study to improve the performance of the deep neural networks, which consists of synthesizing new training samples by applying randomly generated translations, rotations, horizontal flips, zooms, intensity shifts, cutout, and Gaussian noise to the CXR data in the training set to increase data diversity and allow the deep neural networks to learn improved robustness. The proposed deep neural networks were trained using an Adam optimizer with image size of 480$\times$480, batch size of 32, learning rate of 4e-4, 30 epochs, and mean squared error as the loss function. All of the model development was conducted using Python, OpenCV, and the Keras deep learning library with a TensorFlow backend.

\subsection*{Cross-validation and Performance Evaluation}

To evaluate the efficacy of the COVID-Net S deep neural networks developed for computer-aided severity scoring of SARS-CoV-2 lung disease severity, stratified Monte Carlo cross-validation~\cite{Xu} was conducted.  For geographic extent and opacity extent independently, 100 different deep neural networks (50 for geographic extent scoring and 50 for opacity extent scoring) were learned using 100 different random subsets of CXR data from the study (50 for geographic and 50 for opacity).  Each of the 100 different deep neural networks was then tested on 100 different subsets of CXR data that was held out from the learning process.  For each trial, a random subset consisting of 80\% of the CXR data was used to train a deep neural network, with the remaining 20\% of the CXR data held out and used for testing.

To quantify the performance of the deep neural networks learned in this study, we compute the coefficient of determination, R$^2$, between predicted scores outputted by the deep neural networks and scores by expert radiologists for both geographic extent and opacity extent in the test sub-set of CXR data for each trial.  To present a quantitative summary for the cross-validation results, the R$^2$ was averaged over the trials for geographic extent and opacity extent independently, resulting in means and standard deviations across the cross-validation results.

\section*{Results}
\subsection*{Demographic and imaging protocol variables}
Table~\ref{tab:demographic} summarizes the demographic variables and imaging protocol variables of the CXR data used in this study. Note that the majority of the patient cases are from Europe and Asia, and reflects the earlier rise of the COVID-19 pandemic in those two continents.  In addition, the majority of the cases are above the age of 50, with the mean age being 57.5, and is consistent with the greater effect of SARS-CoV-2 on the older population.

\begin{table}[h]
\centering
\caption{Summary of demographic variables and imaging protocol variables of CXR data used in this study. Age, sex, and geographic location statistics are expressed on a patient level, while imaging view and imaging position statistics are expressed on an image level.}
\label{tab:demographic}
\begin{adjustbox}{width=0.4\textwidth}
\begin{tabular}{|c|c|c|}
\hline
\multicolumn{1}{|l|}{\textbf{Age}} & mean $\pm$ std & $57.5 \pm 16.1$ \\ \hline
 & \textless 20 & 1 (0.4\%) \\ \hline
 & 20-29 & 4 (1.5\%) \\ \hline
 & 30-39 & 13 (4.9\%) \\ \hline
 & 40-49 & 20 (7.5\%) \\ \hline
 & 50-59 & 26 (9.7\%) \\ \hline
 & 60-69 & 24 (9.0\%) \\ \hline
 & 70-79 & 29 (10.9\%) \\ \hline
 & 80-89 & 10 (3.7\%) \\ \hline
 & 90+ & 0 (0.0\%) \\ \hline
 & Unknown & 140 (52.4\%) \\ \hline
\multicolumn{3}{|l|}{\textbf{Sex}} \\ \hline
 & Male & 117 (43.8\%) \\ \hline
 & Female & 62 (23.2\%) \\ \hline
 & Unknown & 88 (33\%) \\ \hline
\multicolumn{3}{|l|}{\textbf{Geographic location}} \\ \hline
 & Asia & 29 (10.9\%) \\ \hline
 & North America & 5 (1.9\%) \\ \hline
 & Europe & 196 (73.4\%) \\ \hline
 & Australia & 1 (0.3\%) \\ \hline
 & Unknown & 36 (13.5\%) \\ \hline
\multicolumn{3}{|l|}{\textbf{Imaging view}} \\ \hline
 & PA & 151 (56.6\%) \\ \hline
 & AP & 104 (38.9\%) \\ \hline
 & Unknown & 12 (4.5\%) \\ \hline
\multicolumn{3}{|l|}{\textbf{Imaging position}} \\ \hline
 & Supine & 20 (7.5\%) \\ \hline
 & Upright & 235 (88.0\%) \\ \hline
 & Unknown & 12 (4.5\%) \\ \hline
\end{tabular}%
\end{adjustbox}
\end{table}

\subsection*{Geographic extent and opacity extent analysis at different degrees of severity}
\begin{figure*}[t]
\centering
  \includegraphics[width=1\textwidth]{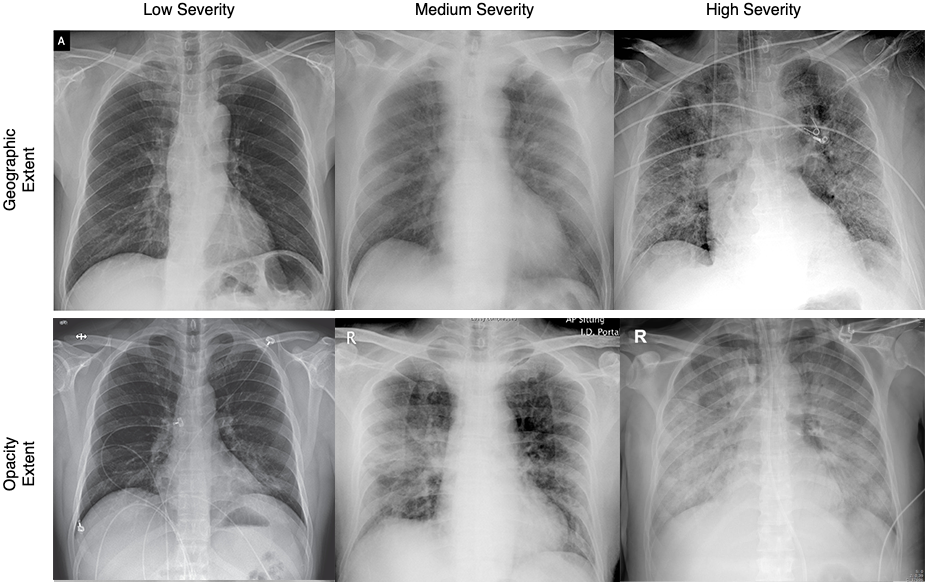}
  \caption{Illustrative SARS-CoV-2 patient cases used in this study with different degrees of geographic extent and opacity extent present in the CXRs. (top row) CXRs exhibiting low, medium, and high geographic extent of lung involvement by ground glass opacity or lung consolidation with respective geographic extent scoring of 1.3, 4.3, and 8.0; (bottom row) CXRs exhibiting low, medium, and high degree of lung opacity with respective opacity extent scoring of 1.0, 4.0, and 6.0.}
  \label{fig:ex-geo-opc}
\end{figure*}

\begin{figure*}[t]
\centering
  \includegraphics[width=1\textwidth]{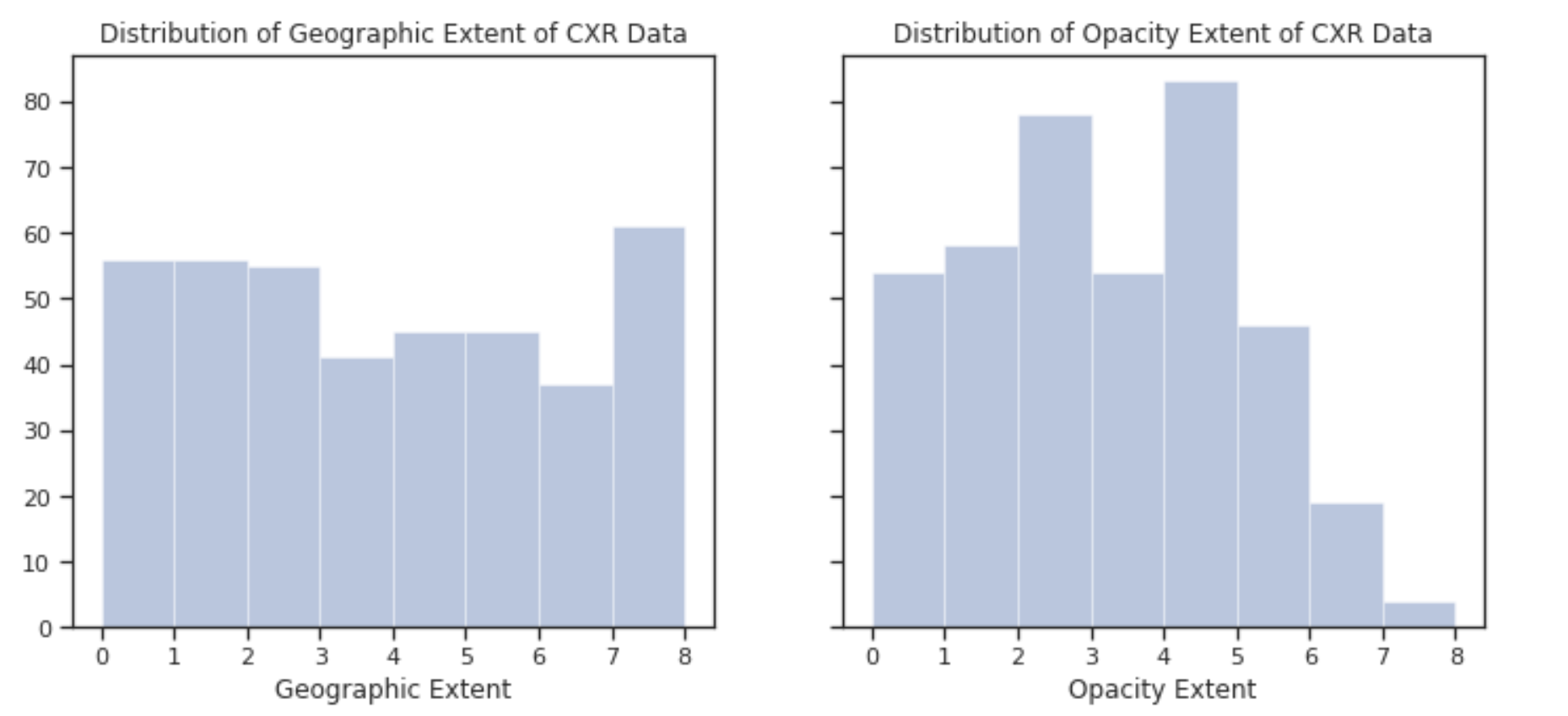}
  \caption{Distribution of geographic and opacity extent scores for patient cases used in this study.}
  \label{fig:ex-geo-opc-hist}
\end{figure*}

Figure~\ref{fig:ex-geo-opc} shows a number of illustrative SARS-CoV-2 patient cases used in this study with different degrees of geographic extent and opacity extent present in the CXRs, with the distribution of geographic and extent scoring for the patient cases shown in Figure~\ref{fig:ex-geo-opc-hist}.  Several observations and insights can be gained from looking at the CXRs of these past SARS-CoV-2 patient cases.  First, it can be observed that the geographic extent of lung involvement by ground glass opacity or lung consolidation have a strong relationship with the degree of lung opacity. In the SARS-CoV-2 patient cases shown here, a visible increase in geographic extent is accompanied by a visible increase in lung opacity.  This relationship between geographic extent and degree of opacity may be useful as a distinguishing property of SARS-CoV-2 infection when reading CXRs.  Second, looking at the SARS-CoV-2 patient cases with low lung severity, it can be observed that the signs of ground glass opacity or lung consolidation can be quite subtle to the human eye, making them difficult to identify visually.  This observation gives insights into the potential challenges involving radiologist readings of patients at very early stages of SARS-CoV-2 infection, given that the extent of ground glass opacity and/or consolidation in the lungs is less prevalent for visual identification.  However, it also brings to light the potential for the use of artificial intelligence for computer-aided decision-making for SARS-CoV-2, with past work~\cite{COVIDNet} demonstrating the ability of deep learning systems to learn and identify such subtle imaging features in CXRs for SARS-CoV-2 detection, and this study assessing the feasibility of deep learning systems for SARS-CoV-2 lung severity scoring.

\begin{table}[]
\centering
\caption{Summary of R$^2$ between predicted scores from the COVID-Net S deep neural networks and the radiologist scores for the 100 experiments (50 deep neural networks for geographic extent scoring and 50 deep neural networks for opacity extent scoring).}
\label{tab:r}
\begin{adjustbox}{width=0.45\textwidth}
\begin{tabular}{|c|c|c|}
\hline
\multicolumn{1}{|c|}{\textbf{}} & \multicolumn{2}{|c|}{\textbf{R$^2$}}\\ \hline \multicolumn{1}{|c|}{} & Geographic extent & Opacity extent\\ \hline
\hline
mean & 0.664  & 0.635 \\ \hline
std & 0.032 &  0.044\\ \hline
\end{tabular}%
\end{adjustbox}
\end{table}

\subsection*{Coefficient of determination analysis}
\begin{figure*}[t]
\centering
  \includegraphics[width=1\textwidth]{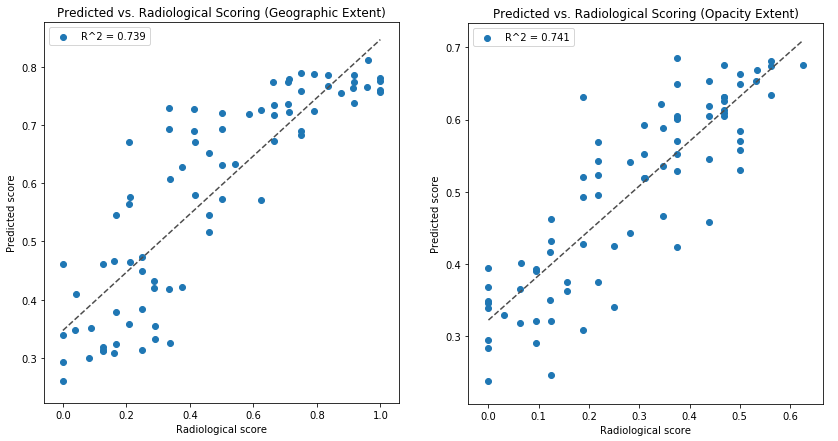}
  \caption{Scatter plots of predicted scores vs. radiologist scores for the best performing networks for geographic extent and opacity extent scoring.}
  \label{fig:scatterplot}
\end{figure*}

Examining the R$^2$ between predicted scores from the COVID-Net S deep neural networks and the radiologist scores for the 100 experiments (50 deep neural networks for geographic extent scoring and 50 deep neural networks for opacity extent scoring) led to number of observations.  First, the deep neural networks yielded R$^2$ of 0.664 $\pm$ 0.032 and 0.635 $\pm$ 0.044 for geographic extent and opacity extent, respectively, in the stratified Monte Carlo cross-validation experiments (see Table~\ref{tab:r}). Second, the best performing networks achieved R$^2$ of 0.739 and 0.741 between predicted scores and radiologist scores for geographic extent and opacity extent, respectively  (see Figure~\ref{fig:scatterplot} for scatter plots of predicted scores vs. radiologist scores for these networks).  Third, the results show that the mean R$^2$ between predicted scores and radiologist scores for geographic extent is higher than that for opacity extent.

\section*{Discussion}

In this study, we hypothesised that computer-aided deep learning algorithms can accurately predict lung disease severity on CXRs associated with SARS-CoV-2 infection against expert chest radiologist ground truths, and the experimental results of study support this hypothesis.  Results from the stratified Monte Carlo cross-validation experiments showed that the learned COVID-Net S deep neural networks could achieve mean R$^2$ between predicted scores and radiologist scores for geographic extent and opacity extent greater than 0.5 when evaluated for 100 different subsets of CXR data from the study (50 for geographic extent scoring and 50 for opacity extent scoring).

Severity scoring for SARS-CoV-2 has gained recent attention due to the rise and continued prevalence of the COVID-19 pandemic across the globe, and the need to assess the severity of a patient who is SARS-CoV-2 positive is crucial for determining the best course of action regarding treatment and care.  Several severity scoring mechanisms have recently been proposed for the severity assessment of SARS-CoV-2.  Wong~\textit{et al.}~\cite{Wong} introduced a scoring scheme for severity quantification of SARS-CoV-2 by adapting and simplifying the Radiographic Assessment of Lung Edema (RALE) score introduced by Warren~\textit{et al.}~\cite{Warren}.  Toussie~\textit{et al.}~\cite{Toussie} introduced a scoring scheme where each lung was divided into three zones (for a total of six zones) and each zone was assigned a binary score based on opacity, with the final severity score being the aggregate of the scores from the different zones.  Borghesi and Maroldi~\cite{Borghesi} introduced a scoring scheme where, similar to Toussie~\textit{et al.}, each lung was divided into three zones, but each zone was instead assigned a score from 0 to 3 based on interstitial and alveolar infiltrates.  Considering the large quantity of patients that are being screened due to the COVID-19 pandemic and the need for expert radiologists to assess the severity of each patient, the use of artificial intelligence for computer-aided severity scoring has strong potential to assist in clinical workflow efficiency given the situation.

This study has a few limitations. First, the data were obtained from various sources and could exhibit bias. Second, disease severity is based on radiologist ground truths, and functional outcomes such as measures of lung function or mortality were not available. Third, the image quality of the CXRs can vary.  Note that although some CXRs have lower resolution, they are observed to be of acceptable diagnostic quality. Fourth and finally, future studies should investigate longitudinal changes in disease severity.

In conclusion, our results support the hypothesis that the use of deep neural networks on CXRs can be an effective tool for computer-aided assessment of lung disease severity, although additional studies are needed before adoption for routine clinical use. This tool may be helpful in ER and ICU settings for triaging patients into general admission or ICU, as well as determining when to put SARS-CoV-2 patients on a mechanical ventilator and when to extubate.

\section*{Acknowledgements}
We would like to thank Natural Sciences and Engineering Research Council of Canada (NSERC), the Canada Research Chairs program, CIFAR, DarwinAI Corp., Nvidia Corp., Hewlett Packard Enterprise Co, and Maya Pavlova.

\section*{Author contributions statement}
Z.Q.L, L.W., A.G.C, A.W., and T.Q.D. conceived the experiment, B.S, A.A., M.H., and T.Q.D. collected the data, B.S, A.A., and M.H. performed the radiological scoring, Z.Q.L., L.W., A.G.C., and A.W. developed the model and prepared the learning procedure, all authors analysed the results. All authors reviewed the manuscript.

\section*{Declaration of interests}
L.W., Z.Q.L., A.G.C. and A.W. are affiliated with DarwinAI Corp.

\section*{Ethics approval}
The study has received ethics clearance from the University of Waterloo (42235). All experimental protocols were approved by University of Waterloo. All methods were carried out in accordance with University of Waterloo ethics guidelines and regulations. Informed consent was obtained from all participants.

\bibliography{sample}

\end{document}